\def\erg{{\rm\thinspace erg}}

\def\keV{{\rm\thinspace keV}}
\def\km{{\rm\thinspace km}}
\def\kpc{{\rm\thinspace kpc}}

\def\Msun{\hbox{$\rm\thinspace M_{\odot}$}}

\def\s{{\rm\thinspace s}}
\def\yr{{\rm\thinspace yr}}
\def\Gyr{{\rm\thinspace Gyr}}

\def\ergps{\hbox{$\erg\s^{-1}\,$}}

\def\kmps{\hbox{$\km\s^{-1}\,$}}

\def\Msunpyr{\hbox{$\Msun\yr^{-1}\,$}}

\documentclass[usegraphicx]{mn2e}
\usepackage{amsmath}
\begin{document}

\title[Sound Waves in Perseus]{Do sound waves transport the AGN
  energy in the Perseus Cluster?
  } \author[A.C. Fabian et al] {\parbox[]{6.5in}{{
A.C.~Fabian$^1\thanks{E-mail: acf@ast.cam.ac.uk}$, S.A. Walker$^1$,
H.R. Russell$^1$, C. Pinto$^1$, J.S.~Sanders$^2$ and C.S.~Reynolds$^{3}$ }\\
\footnotesize 
$^1$ Institute of Astronomy, Madingley Road, Cambridge CB3 0HA\\
$^2$ Max-Planck-Institut f\"ur Extraterrestrische Physik,
Giessenbackstrasse 1, D-85748 Garching, Germany \\
$^2$ Dept. of Astronomy, University of Maryland, College
Park, MD 20712-2421, USA\\
  }}

\maketitle
  
\begin{abstract}
The level of random motions in the intracluster gas lying between 20
and 60 kpc radius in the core of the Perseus cluster has been measured
by the Hitomi Soft X-ray Spectrometer at $164\pm10\kmps$. The maximum
energy density in turbulent motions on that scale is therefore low. If
dissipated as heat the turbulent energy will be radiated away in less
than 80~Myr and cannot spread across the core. A higher velocity is
needed to prevent a cooling collapse. Gravity waves are shown to
travel too slowly in a radial direction. Here we investigate
propagation of energy by sound waves. The energy travels at
$\sim1000\kmps$ and can cross the core in a cooling time. We show that
the displacement velocity amplitude of the gas required to carry the
power is consistent with the Hitomi result and that the inferred
density and temperature variations are consistent with Chandra
observations.
\end{abstract}


\section{Introduction}
The Soft X-ray Spectrometer on the Hitomi satellite has recently
measured the level of gas motions in the X-ray bright IntraCluster
Medium (ICM) in the core of the Perseus cluster to an unprecedented
precision of $10 \kmps$ (Hitomi collaboration 2016). Small-scale
motions ($<10\kpc$) in a region 20 to 60 kpc from the central Active
Galactic Nucleus (AGN) are $164\pm10 \kmps$; larger scale shear across
that region is also low, $\sim150\pm70\kmps$.  The energy density of
turbulent motions is at most 4 per cent of the thermal energy density
of the hot intracluster gas.

In the absence of a heat source the hot gas in this region would cool
radiatively, due to the emission of X-rays, on a timescale of
$1-2 \Gyr$ forming a massive cooling flow of several hundred
$\Msunpyr$ (see e.g. Fabian 2012). Although the temperature of the hot
gas does drop towards the centre, little gas is seen below 3 keV,
ruling out such a cooling flow.  If all the energy in random motions
is ascribed to turbulence which is then dissipated as heat, then it
will be used up in 4 per cent of the cooling time: it needs to be
replenished on the same timescale, which is about
$8\times10^7\yr$. However if the velocity at which energy flows is
limited to $164\kmps$ then the energy can only travel 13~kpc in that
timescale, before being lost as radiation.  Energy needs to be
transported from the energy source -- the AGN -- at a much higher
velocity of at least $700 \kmps$, close to the speed of sound in the
region ($\sim1000 \kmps$).  One energy transport mechanism that can do
this is sound waves. The matter does not flow but the energy
does. Here we explore further the case for the importance of sound
waves in the Perseus and other cool core clusters. The observed
velocity dispersion is consistent with that required to transport the
energy required for heating the cluster core.

Many studies of cool-core clusters (B\^irzan et al (2004, 2012); Dunn
\& Fabian (2006, 2008); McNamara \& Nulsen 2012; Fabian 2012) have
shown that the bubbles produced by jets from the central black hole
represent a considerable energy flow which is capable of balancing
radiative cooling (e.g. Churazov et al 2002). What has remained unclear
is the means by which the energy from the bubbles is transported
throughout the core and distributed in a roughly isotropic
manner. Ripples with a wavelength $\sim15\kpc$ were seen in the first
deep Chandra images of the Perseus cluster and interpreted as sound
waves (Fabian et al 2003). Their amplitude was consistent with the required
energy flux. Issues of dissipation were explored by Fabian et al
(2005). Sound wave propagation and dissipation in intracluster gas
were simulated by Ruszkowski et al (2004a,b) and Sijacki et al (2005)
and have been further discussed by Mathews, Faltenbacher \& Brighenti
(2006) and by Heinz \& Churazov (2005).

The ripples became clearer in yet deeper Chandra observations (Fabian
et al 2006; Sanders \& Fabian 2007). The expected temperature changes
associated with sound waves were at the limit of detectability in
these observations and were not seen. Later work on surface brightness
fluctuations found in these observations were interpreted by
Zhuravleva et al (2014) as due to turbulence. This was then extended
to cover 2 X-ray colour bands, making the work sensitive to
temperature fluctuations and with a goal of testing the the equation
of state of the fluctuations.  The ripples were not however bright
enough to be separately analysed (see Zhuravleva et al 2016 and
Section 3). A similar analysis has been made of the ICM around M87 by
Ar\'evalo et al (2016).

\section{The generation and propagation of turbulent motions}

Reynolds, Balbus \& Schekochihin (2015) have examined the generation
of subsonic turbulence through 3D hydrodynamical simulation of
AGN-like events. Gravity waves (g-modes) are launched by these events
into the ICM and decay to volume-filling turbulence. They find however
that the process is very inefficient with less than 1 per cent of the
energy injected by the AGN activity ending up as turbulence in the
ICM.  Yang \& Reynolds (2016) have then used further hydrodynamical
simulations to explore momentum-driven jet feedback, confirming that the
level of turbulent heating is low, accounting for only about one per
cent of the energy.  Turbulence is not necessarily an essential
ingredient of cluster AGN feedback.

\begin{figure}
  \centering
  \includegraphics[width=0.99\columnwidth,angle=0]{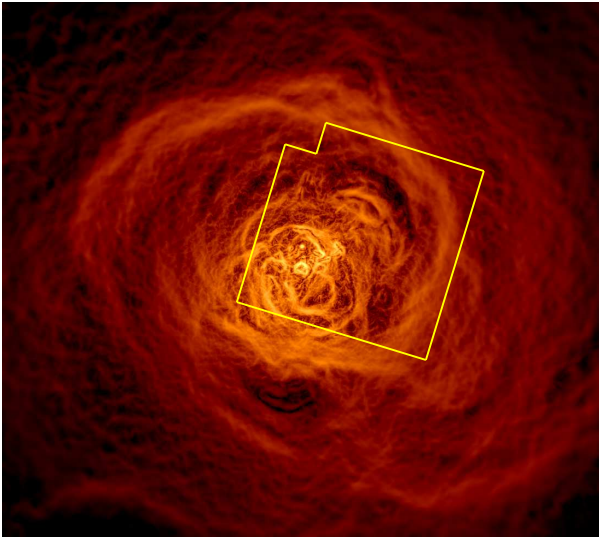}
  \caption{The core of the Perseus cluster filtered with an
    edge-detection algorithm (Sanders et al 2016b) and overlaid with the
    field of view of the Hitomi SXS (3x3 arcmin, or about 60x60~kpc).   }
\end{figure}

\begin{figure}
  \centering
  \includegraphics[width=0.99\columnwidth,angle=0]{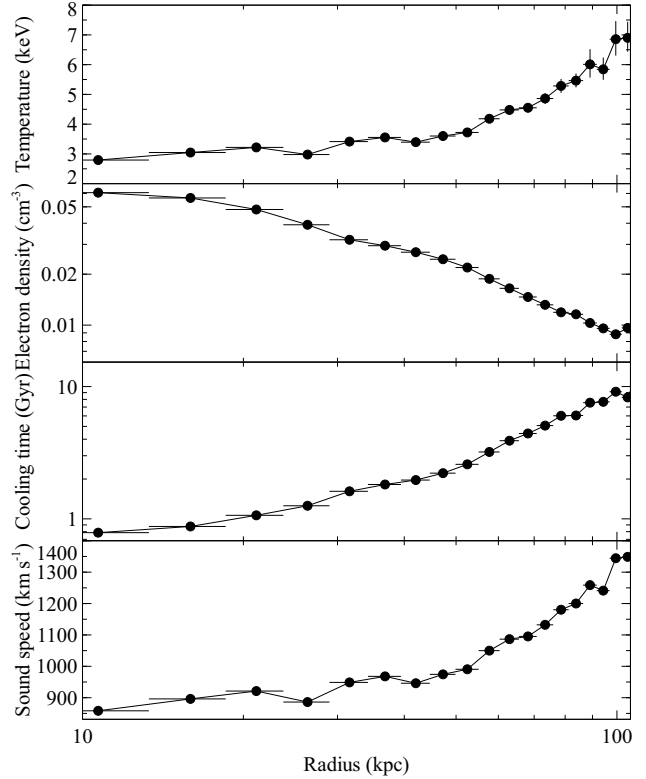}
  \caption{Profiles of the intracluster temperature, density,
    radiative cooling time and sound speed in the Perseus cluster.}
\end{figure}

Still, it is possible that processes not captured by these simple
hydrodynamic models might lead to significantly stronger AGN-driving
of $g$-modes (see Section~6 of Reynolds et al. 2015) and so it is
interesting to examine the subsequent propagation speed of $g$-modes.
Working within a local/WKB approximation, the dispersion relation for
a particular $g$-mode with (angular) frequency $\omega$ and wavevector
${\bf k}$ is $\omega^2=N^2k_h^2/|{\bf k}|^2$, where $k_h$ is the
horizontal component of the wavevector and $N$ is the
Brunt-V\"ais\"al\"a frequency given by
\begin{equation}\label{eq:bv} N^2=-\frac{c_s^2}{\gamma}\frac{\partial
(\ln\rho)}{\partial z}\frac{\partial (\ln P\rho^{-\gamma})}{\partial
z},
\end{equation} ($c_s$ is the adiabatic sound speed, $\rho$ is density,
$P$ is pressure, and $\gamma$ is the adiabatic index which we take to
be $\gamma=5/3$). Energy propagates at the group velocity, ${\bf
v}_g=\nabla_{\bf k}\omega$, and so the outwards/radial speed of
propagation is
\begin{equation} v_{g,r}=\frac{\partial \omega}{\partial
k_r}=\sin\theta\cos^2\theta\frac{N}{k_r}.
\end{equation} Here, $\theta$ is the angle that the propagation
direction makes with the outward pointing direction and $k_r$ is the
radial component of the wavevector.  Since
$|\sin\theta\cos^2\theta|<2/3\sqrt{3}\approx 0.38$, we see that
$v_{g,r}<0.38(N/2\pi)(2\pi/k_r)$ --- this has the simple
interpretation that, with each buoyant cycle, the $g$-mode propagates
upwards by (at most) half of a radial wavelength $\lambda_r=2\pi/k_r$.

If $H_\rho$ and $H_S$ are the density and entropy scale-heights,
equation~(1) becomes $N=c_s/\sqrt{\gamma H_\rho H_S}$ and we
find that
\begin{equation} v_{g,z}< 0.047c_s\frac{\lambda_r}{\sqrt{H_\rho H_S}},
\end{equation} so we immediately see that $g$-modes propagate very
subsonically provided that $\lambda_r<H_\rho,H_S$ (note that, if this
condition were violated, the local approximation used here would break
down).  Measurements of the temperature and density profiles in
Perseus (Fig.~2) imply scale-heights of $H_\rho,H_S\sim 30-50$\,kpc
which, when compared with the $\sim 15$\,kpc radial wavelength of the
observed ripples, yields radial propagation speeds for the $g$-modes
of $v_{g,z}<0.02c_s\approx 20-30$\,km\,s$^{-1}$. Even taking the
maximum wavelength possible ($\lambda_r\sim H_\rho$) yields
$v_{g,z}<0.05c_s\approx 50-70$\,km\,s$^{-1}$.  These $g$-mode
propagation velocities are at least an order of magnitude too small to
replenish the decaying turbulence, if the turbulence is indeed
responsible for offsetting the radiative cooling.

The sharp high-energy edge to the Fe He$\alpha$ resonance line
observed from the Perseus cluster by the Hitomi SXS  leaves no room 
for any significant component with larger velocity motions.

\section{Sound propagation}

The energy in a sound wave is carried at the speed of sound, as the
particles oscillate about their mean position at a much lower
(displacement) velocity.  If the energy density due to random motions
is $\frac{1}{2} \rho v_{\rm r}^2$ where $v_{\rm r}$ is the gas velocity then the
maximum rate at which energy flows from a central source is
\begin{equation} 
{\cal P_{\rm v}}=2 \pi r^2 \rho v_{\rm r}^3.
\end{equation} The energy flux due to sound waves which have a
displacement velocity amplitude $v_{\rm s}$ is
\begin{equation} 
{\cal P_{\rm s}}=4 \pi r^2 c_{\rm s}\rho v_{\rm s}^2.
\end{equation} Sound waves can therefore transport $2 c_{\rm s}/v_{\rm r}$
times more power than random motions alone, for a given measured velocity.  
This is over an order of magnitude increase in the case of the Perseus
cluster considered here, where $c_{\rm s}\sim 1000 \kmps$ and $v_{\rm r}\sim
164 \kmps$.

In order to determine the displacement velocity from the observed
broadening of emission lines we need to correct for projection
effects. We achieve that by simulating a cluster as many cubic cells,
of size matching the SXS 10 kpc pixels, each containing gas of the
appropriate density and temperature to match observed profiles
(Fig.~2). Assigning a radial velocity to the gas in each cell
according to a radial sine wave enables us to assemble the
emission-weighted line profile as a function of projected radius
(Fig.~3: the intrinsic profile is shown in the upper panel and the
profile which would be observed, after convolution with the
instrumental and thermal broadening, is shown in the lower panel). The
sharp peak is due to the brightest gas along any line of sight lying
in the plane of the Sky where the apparent radial displacement
velocity of the gas is zero.  We measure one standard deviation of
that profile (68 per cent of the total emission from the line centre)
and use that value. It varies with radius and is between about one
third to one half of the amplitude of the true displacement
velocity. This yields another factor of at least four in favour of
sound wave propagation. (The $200\kmps$ used for Fig. 3 is a fiducial
value.)

If all the measured dispersion is attributed to radial sound waves
then their intrinsic velocity amplitude would be about $300\kmps$. It is
possible, of course, that the observed dispersion is a combination of
sound waves of lesser amplitude with other random motions and
turbulence.

Future microcalorimeters with higher spectral resolution, such as the
planned X-ray Integral Field Unit (X-IFU: Barret et al 2016) on ESA's
Athena X-ray Observatory may be able to measure the non-gaussian
nature of the perturbations caused by sound waves.

We note that the recent simulations of Yang \& Reynolds (2016) provide
strong support for the idea that sound waves are a major
contributor to the ICM velocities.  They use a Helmholtz
decomposition to separate compressible motions (sound waves and
shocks) from incompressible motions (turbulence and large scale
circulation) in their simulations of a Perseus-like cluster with
self-regulation by a jetted AGN. Throughout most of the volume of the
cluster core (i.e. the region away from the ``jet cones’’) at a
typical snapshot in time, the compressible motions dominate with a
magnitude $v_{\rm s}\sim 150\kmps$; turbulent velocities are somewhat
less ($v_{\rm turb}\sim 100\kmps$) with bulk flow and circulatory
motions being smaller yet (H.-Y.~Karen~Yang, private communication).
Since the Yang \& Reynolds simulations adopt an ideal hydrodynamic
framework, there are no mechanisms that can efficiently dissipate the
sound waves, and thus the heating actually occurs via bubble mixing
and weak shocks in the jet-cone, with a gentle circulation that cycles
a large part of the ICM through this region on Gyr timescales.  The
expectation is that, given the inclusion of a suitable dissipation
mechanism, sound wave heating will be important in future developments
of these models.

\begin{figure}
  \centering
  \includegraphics[width=0.99\columnwidth,angle=0]{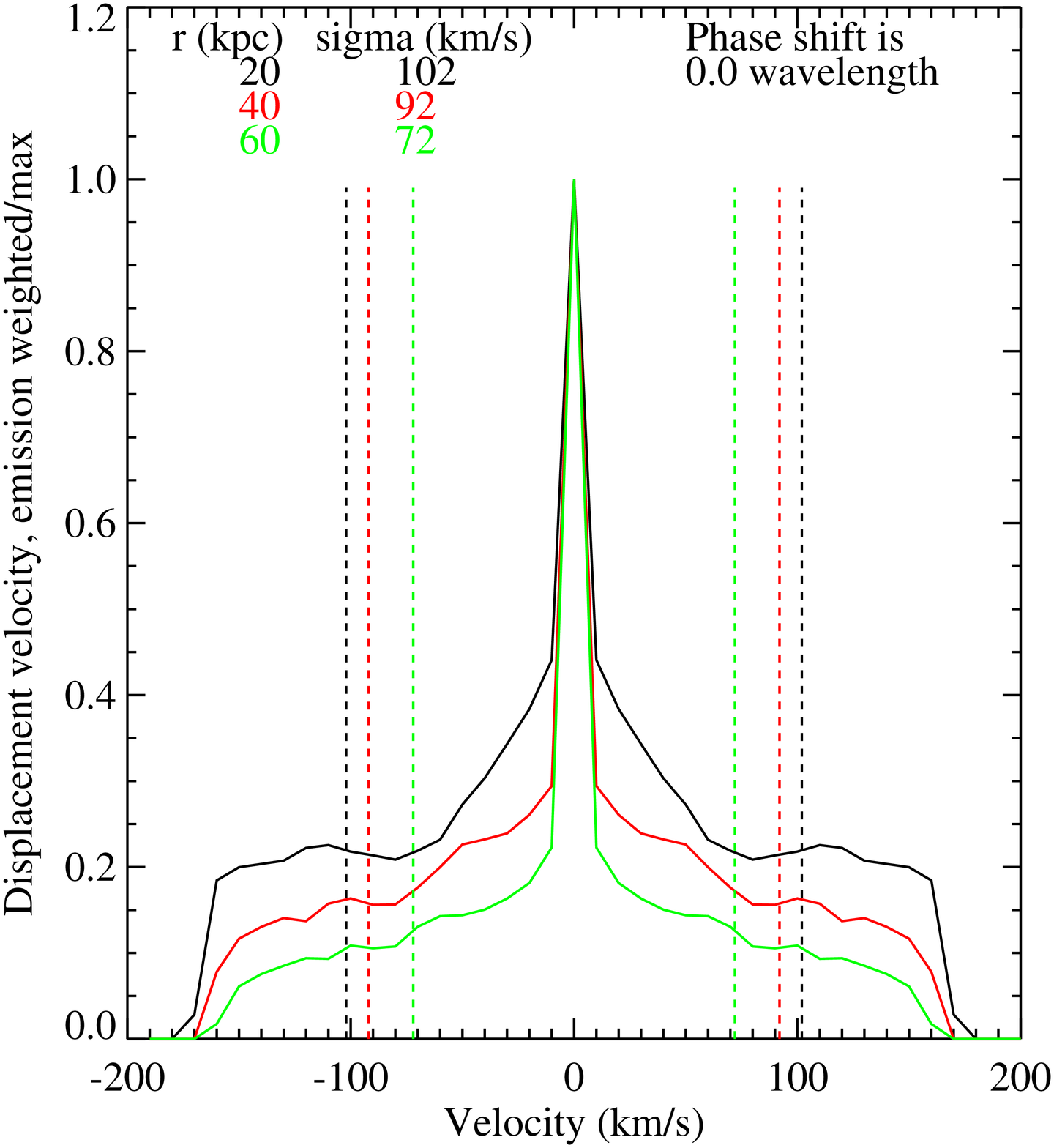}
\includegraphics[width=0.99\columnwidth,angle=0]{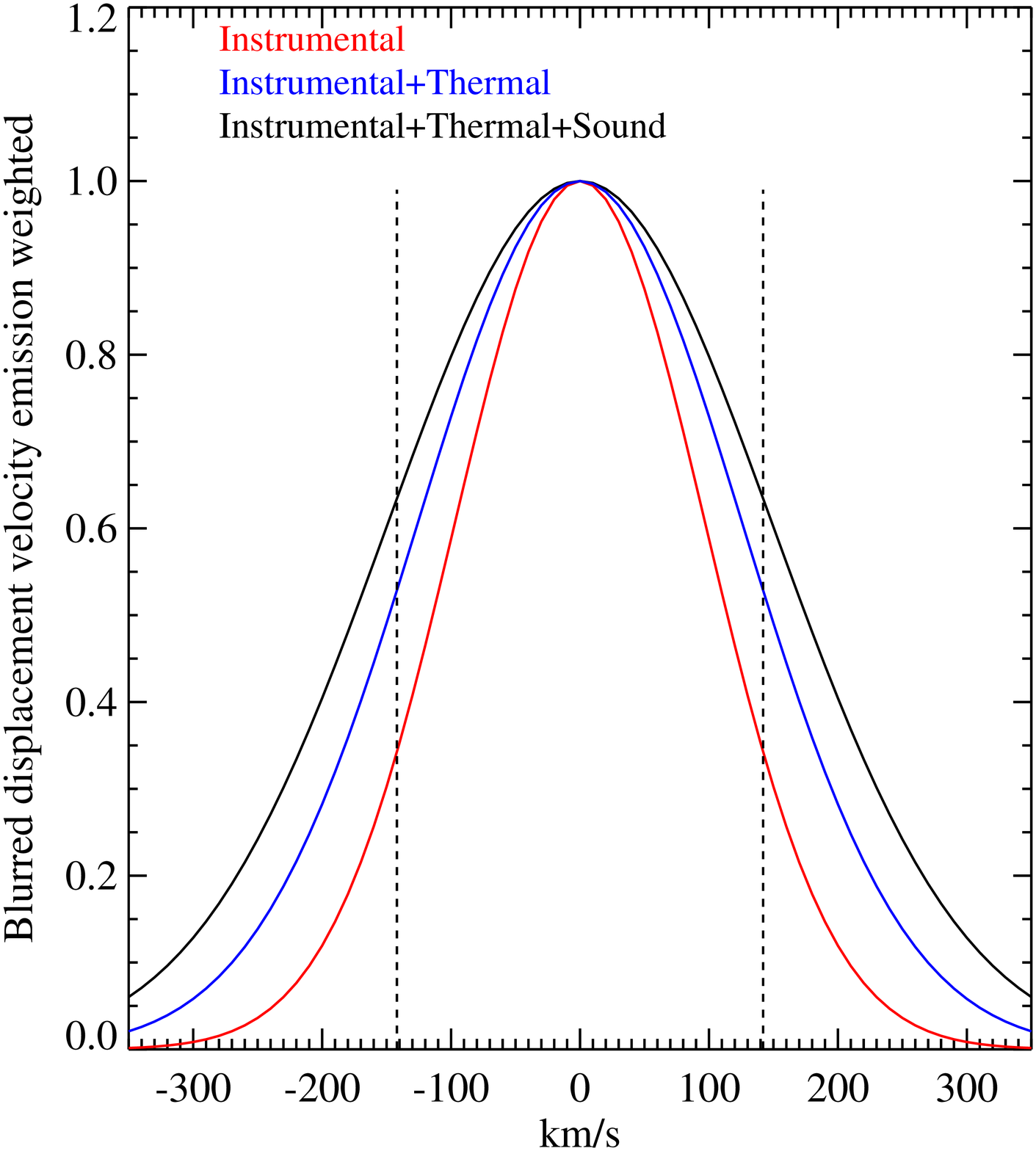}
\caption{Top: Simulated line profiles due to the effect of the
  displacement velocity of sound waves in the ICM of the Perseus
  cluster. The profiles shown are for three different lines of sight
  passing at increasing distances from the core, spanning the range
  covered by the Hitomi observation (20, 40 and 60 kpc), and using the
  same pixel size as the Hitomi SXS. The line of sight component of
  the displacement velocity was integrated through the cluster along
  these sight lines, weighted by the emission measure of the gas. The
  corresponding coloured dashed vertical lines show the width
  containing 68 percent of the line strength. There is only a small
  change if the phase of the sound waves is shifted by half a
  wavelength. Bottom: Appearance of line profile expected at 40 kpc
  offset, after convolution with both instrumental and thermal
  broadening.  }
\end{figure}

\section{Temperature fluctuations in sound waves}

Adiabatic sound waves are accompanied by temperature fluctuations
\begin{equation} 
\frac{\Delta T}{T}=\frac{2}{3}\frac{\Delta
n}{n}\approx\frac{1}{3}\frac{\Delta I}{I}.
\end{equation} where $I$ is the soft X-ray surface emissivity.

Projection effects reduce the apparent amplitude of sound waves as
observed in X-rays in intracluster gas (Graham et al 2008), such that
waves with a density amplitude of 5 per cent, which leads to an
emissivity fluctuation of about 10 per cent (ie $\propto n^2$) can
appears as a surface brightness fluctuation of only 3--5 per
cent. (The observed fluctuations are about $\pm3$~per cent at radii of
20-60~kpc in the Perseus cluster; Sanders \& Fabian 2007).

Projection also affects the amplitude of the accompanying temperature
fluctuations: all lines of sight contain variations, but some have
more than others.  In order to estimate these affects we have used the
same simulation as mentioned above for the velocity only now we
calculate the integrated spectrum along the line of sight. (The pixels
and response files now match those of Chandra.) Such spectra are then
converted to hardness ratios (Fig.~4). We employ the bands $0.5-4\keV$
and $4-8\keV$ as used by Zhuravleva et al (2016); the harder band is
much more temperature dependent than the softer one.  It is clear that
it is very difficult to measure periodic variations in hardness ratio
unless the exposure is significantly longer that the current one. We
have also measured from the simulation the hardness ratio of the
relative fluctuations in each band, which Zhuravleva et al (2016) and
Churazov et al (2016) show is sensitive to the effective equation of
state of the ICM. The resulting value is $1\pm0.2$ which is consistent
with isolated isothermal perturbations. The adiabatic perturbations
used in the simulation appear isothermal due to the large effect of
projection.

\begin{figure}
  \centering
  \includegraphics[width=0.99\columnwidth,angle=0]{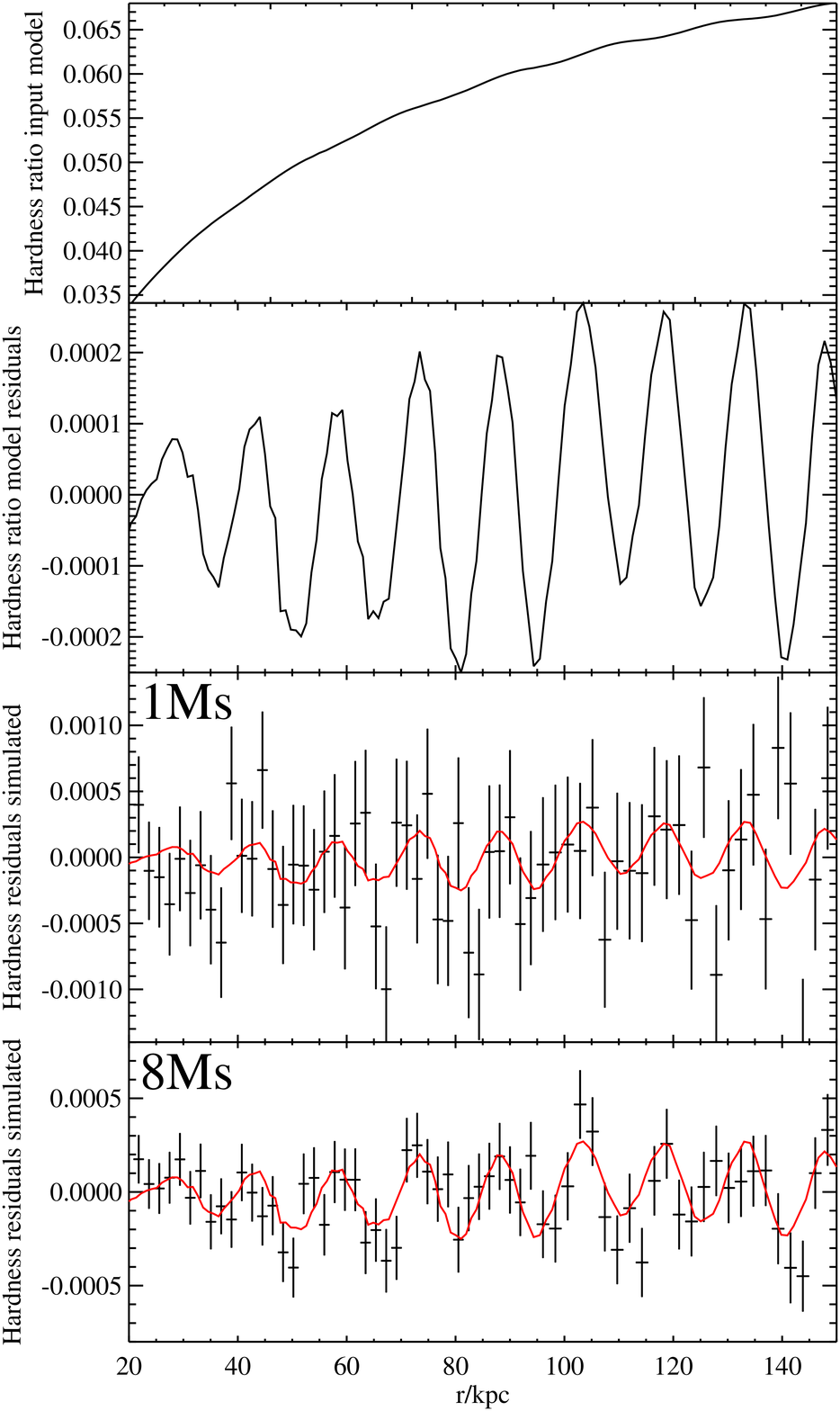}
  \caption{Simulated hardness ratio profiles, using 4.0-8.0
    keV for the hard band and 0.5-4.0 keV for the soft band. Top: The
    input model for the hardness profile, taking into account
    projection effects. Second: The residuals in the input hardness
    ratio model due to the input sound waves. Third: Hardness ratio
    residuals in a 1Ms simulated observation, compared to the input
    model (red). Fourth: Hardness ratio residuals in an 8Ms simulated
    observation. }
\end{figure}

\section{Generation and Dissipation of sound}

We are here considering the situation where the bubbling energy of the
AGN is converted into mechanical motions in the gas which is 
transported throughout the cluster core and dissipated there. The
above estimates and the observed limiting level on turbulent velocity
preclude turbulence as the sole basis of the mechanical motions which heat
the whole cluster core. Sound waves remain possible as the transport
mechanism because they transport energy  much faster. 

Sound waves can be generated during the inflation stage of the
bubbles, particularly if the jets are unsteady. Malzac (2014) has
successfully modelled the jets of some X-ray binaries by relating the
spectrum of variations in the accretion flow to the acceleration of
matter in the jet. The AGN at the centre of the Perseus cluster, in
the galaxy NGC1275, is known to vary by over a factor of 30 over the
past 4 decades (Fabian et al 2015 and references therein). It may plausibly  
vary by even larger factors and many times over the much longer time 
taken to form the bubbles. 

The bubbles may themselves oscillate especially as they detach and
rise. The natural frequency of a bubble, the Minnaert frequency 
\begin{equation}
f_{\rm M}=\frac{1}{2\pi r_{\rm b}}\left(\frac{3\gamma p_{\rm
      A}}{\rho}\right)^{1/2}.
\end{equation}
This leads to a wavelength of about $4r_{\rm b}$, where $r_{\rm b}$ is
the bubble radius, $p_{\rm A}$ is the ambient pressure and $\rho$ the
density of the surrounding medium. Such a large wavelength
($\sim 30\kpc$) is probably too large to be relevant.

It is also likely that the $5\times 10^{10}\Msun$ of cold molecular
gas in the core of the cluster (Salom\'e at al 2011) oscillates in
response to the bubbling process and so generates further sound waves.

The power of the sound,
\begin{equation}
  \begin{split}
    {\cal P_{\rm s}}&=4 \pi r^2 c_{\rm s} \rho v_{\rm s}^2\\
    &= 2.9\times 10^{44} \left(\frac{r}{40\kpc}\right)^2
    \left(\frac{v}{200\kmps}
    \right)^2\ergps
    \\
    &=4 \pi r^2 \frac{\delta P^2}{\rho c_{\rm s}},
\end{split}
\end{equation}
where P is the gas pressure and $\delta P$ is the pressure change
associated with the sound wave. 
Note that
\begin{equation}
\frac{\delta P}{P}=\frac{v}{c_{\rm s}}. 
\end{equation}

The power scaling is close to that required to compensate for
radiative energy losses in the cluster core (see Fig.5 of Sanders \&
Fabian 2007). The exact level required depends on the length of time
since the present configuration of intracluster gas has lasted. Given
the level of sloshing seen in the cluster this could be 4-5~Gyr, which
agrees with the required power estimates of
$2.5-5\times 10^{44}\ergps$ at 40~kpc radius. The power required at 10
kpc is $6-8\times 10^{44}\ergps$, with the difference having been
dissipated at radii between 10 and 40~kpc. 

The required dissipation length is 50--70\,kpc.  How the energy might
be dissipated by sound waves has been considered by Ruszkowski et al
(2004a) and Fabian et al (2005) and remains an open question. Simple
hydrodynamic viscosity and conduction may be insufficient. Tangled
magnetic fields may give a bulk viscosity as sound wave propagation
changes the field configuration in a non-reversible manner.
 
Sound waves will refract and reflect from density changes and
discontinuities, including cold fronts, meaning that the situation
will not be simple but, at radii of a few wavelengths, approach that
of a noisy room.

Weak shocks and possible sound waves have previously been reported
from M87 (Forman et al 2007), the Centaurus cluster (Sanders et al
2009, 2016a) and A2052 (Blanton et al 2011). Difficulties in detecting
sound waves, in part due to projection effects, have been explored by
Graham et al (2008) and reveal that most current exposures are far too
short. Walker et al (2016) have shown that edge-detection (Gaussian
Gradient Magnitude GGM) techniques can increase the contrast of sound
waves in X-ray images. That work and the GGM images of Sanders et al
(2016b; see also Fig.~1 here) show that elongated, azimuthal
structures are relatively common in cool core clusters. 

The Hitomi SXS result demonstrates that cluster AGN feedback cannot
just involve turbulence. We show here that sound waves, first observed
with the necessary level of surface brightness
fluctuations in 2003, have a velocity amplitude consistent with
both the Hitomi spectrum and the propagation of AGN power through the
cool core of the Perseus cluster. We anticipate that generation of
powerful sound waves by AGN feedback is present in all cool core
groups and clusters and is a major part of the feedback loop.
  
\section*{Acknowledgements}
ACF, CP, CSR and HRR thank the Hitomi collaboration for the
opportunity to participate in the analysis of the SXS data.  ACF, CP,
HRR and SAW acknowledge support from ERC Advanced Grant FEEDBACK,
340442.



\end{document}